\documentstyle[twocolumn,aps,psfig,amssymb]{revtex}

\begin{document}
\title{Real--time Kadanoff--Baym approach to plasma oscillations in
a correlated electron gas}

\author{N.-H. Kwong\footnote{Cooperative Excitation Project, ERATO, Japan
Sciences and Technology Corporation, Optical Sciences Center,
University of Arizona,
Tucson, AZ 85721, USA} and M. Bonitz}
\address{Fachbereich Physik, Universit{\"a}t Rostock\\
Universit{\"a}tsplatz 3, D-18051 Rostock, Germany}
\date{\today}
\maketitle
\begin{abstract}
A nonequilibrium Green's functions approach to the collective response of
correlated  Coulomb systems at finite temperature
is presented. It is shown that solving
 Kadanoff--Baym type equations of motion for the two--time correlation
 functions with the external perturbing field included
allows to compute the plasmon spectrum with collision effects in a
systematic and consistent way. The scheme has ``built-in''
sum rule preservation and is simpler to
implement numerically than the equivalent
equilibrium approach based on the Bethe-Salpeter equation.
\end{abstract}
\pacs{  }

The dynamic properties and the plasmon spectrum of Coulomb systems
continue to attract the interest of researchers in many fields, in particular
condensed matter theory e.g.
\cite{schoene-etal.98,schindlmayr-etal.98,tamme-etal.99,green-etal.85},
plasma physics e.g. \cite{theobald-etal.96,wierling-etal.,green-book} and
electronic bilayer liquids \cite{kalman-etal.99}.
This is due to the fact that the density
response  to an external perturbation, given e.g. by the dynamic structure factor
$S(q,\omega)$,  is a sensitive indicator of the state of a charged particle system
which can be directly measured in x-ray or light scattering and
electron-energy-loss experiments. This is
particularly valuable for strongly coupled many-body systems, such as dense
plasmas, metals or semiconductors at low temperature.

In recent years there has been considerable progress in the theoretical account
of the impact of correlations
among the carriers on collective excitations, i.e. in theories which go beyond
the mean--field level (time-dependent Hartree/Vlasov or
random phase approximation, RPA).
It is now commonly accepted that any model has to obey certain consistency
requirements which, in particular, are related to the preservation of sum rules
for the inverse dielectric function, see Refs.
\cite{schoene-etal.98,schindlmayr-etal.98,tamme-etal.99,green-etal.85}
for a discussion. Among the successful approaches, we mention attempts to
construct local field corrections (see references in \cite{green-etal.85}), kinetic
theory concepts to incorporate collisions into the dielectric function, e.g.
\cite{mermin70}, and Green's functions methods originally
developed by Baym and Kadanoff \cite{baym-etal.61} and others.
The latter approach is of particular interest as it allows for a systematic
first principles
treatment of correlations, and sum rule preservation is easily guaranteed
by using so-called
{\em conserving} approximations for the Green's functions
\cite{baym-etal.61,green-etal.85,green-etal.87}.

In most Green's functions treatments
\cite{baym-etal.61,schoene-etal.98,schindlmayr-etal.98,tamme-etal.99,green-etal.85,green-etal.87},
linear response theory is used which
relates the density response function to the retarded
{\em one-particle-one-hole Green's function of the unperturbed system}
the calculation of which is the central problem.
This task is accomplished by
solving a Bethe-Salpeter equation (BSE), the quality of the results
being determined by the choice of the four-point particle-hole-irreducible
(PHI) vertex $K$. While high-level approximations for $K$ have been investigated for
metals at zero temperature \cite{green-etal.85,green-etal.87},
{\em finite-temperature} treatments are restricted to much simpler approximations
\cite{schindlmayr-etal.98} and have to
neglect vertex corrections which are important for sum rule preservation
\cite{schoene-etal.98,tamme-etal.99}. Moreover, the collective response
from a {\em nonequilibrium state}, which is of high interest e.g. in laser
excited semiconductors or laser plasmas, is completely out of reach in the
BSE approach.

In this letter, we present a scheme which allows i) to compute the linear
response at finite temperature fully including vertex corrections, ii) the
nonlinear response to a strong perturbation and iii) the response from
an arbitrary nonequilibrium state.
Our approach is based on directly computing the
time-dependent density fluctuations of the electron gas under an
external perturbing field, from which we obtain the density response
function. We calculate the {\em nonequilibrium} (two-time)
{\em one-particle}
Green's functions by solving a generalized Kadanoff-Baym type equation with the
external perturbing field included.
The density fluctuation is given by the equal-time limit of the linearized
deviation of the one-particle Green's function from its unperturbed value.
If the unperturbed system is in equilibrium,
sum rules are again preserved by adopting conserving approximations
of the self energy \cite{baym-etal.61}.

There is a one-to-one equivalence between levels of approximations
in the two approaches.  For each choice of the approximate self energy
in the present approach, its formal
functional derivative with respect to the one-particle Green's function
gives the equivalent PHI vertex in the
Bethe-Salpeter approach.  However,
in investigations involving more sophisticated approximations,
our approach has the advantage that
the self-energies are formally much simpler, and hence easier
to implement numerically, than their
equivalent PHI vertices.
Moreover, our approach can be extended to compute the
nonlinear response of a correlated electron gas. We underline that
this efficient and conserving calculational scheme is not limited to the
problem of plasma oscillations but is of interest also for current studies of
finite temperature spin modes in Fermi liquids \cite{farinas-etal.99} and
finite temperature collective excitations in nuclei \cite{dang-etal.98}.
In the following,
to avoid confusion in terminology, we append the suffix ``-BS'' to the
labels of the approximations in the Bethe-Salpeter approach.

We consider a correlated electron gas in a neutralizing background
under the influence of an external potential $U$ described by the
hamiltonian ${\hat H} = {\hat H}_{\rm sys}+{\hat H}_{\rm ext}$, with
\begin{eqnarray}
{\hat H}_{\rm sys}&=&\sum\limits_{{\bf k}}\epsilon_{{\bf k}}
  a^{\dagger}_{{\bf k}} a_{{\bf k}} +
\frac{1}{2}\sum\limits_{{\bf k}_1 {\bf k}_2, {\bf q}\ne 0} V(q)
  a^{\dagger}_{{\bf k}_1+{\bf q}} a^{\dagger}_{{\bf k}_2-{\bf q}}
  a_{{\bf k}_2} a_{{\bf k}_1},
\nonumber\\
{\hat H}_{\rm ext}&=&\sum\limits_{{\bf q}} U(-{\bf q},t)\sum\limits_{{\bf k}}
  a^{\dagger}_{{\bf k+q}} a_{\bf k}.
\label{hsys}
\end{eqnarray}
Here $k,q$ are momenta, $\epsilon_k$ is the one-particle energy, and
$V(q)$ and $U(q)$ are the spatial
Fourier components of the Coulomb potential and the external potential,
respectively. $a^{\dagger}$
($a$) are Heisenberg creation (annihilation) operators evolving with the
total hamiltonian ${\hat H}$. (Spin degrees of freedom are not of interest
for our analysis, so spin indices will be suppressed.)
The nonequilibrium properties of the inhomogeneous electron gas are
defined by  the two-time correlation functions
\begin{eqnarray}
G^<({\bf k+q},t_1;{\bf k},t_2)&=&
  i\langle a^{\dagger}_{{\bf k}}(t_2) a_{{\bf k+q}}(t_1)\rangle;
\nonumber\\
G^>({\bf k+q},t_1;{\bf k},t_2)&=&
  -i\langle  a_{{\bf k+q}}(t_1)a^{\dagger}_{{\bf k}}(t_2) \rangle,
\label{ggl}
\end{eqnarray}
where the statistical averaging is over the density operator of the
initial state.
In particular, the density is given by
$n({\bf q},t)=-i \sum_{{\bf k}} G^<({\bf k+q},t;{\bf k},t)$. The time
evolution
of $G^>$ and $G^<$ is governed by the Kadanoff-Baym equations (KBE)
\cite{baym-etal.61},
\begin{eqnarray}
\left( i\hbar \frac{\partial}{\partial t_1} -
\epsilon_{{\bf k}_1} \right)G^{\gtrless}({\bf k}_1t_1; {\bf k}_2t_2) \qquad
\nonumber\\
=
\sum\limits_{{\bf q}} U(-{\bf q},t_1)
G^{\gtrless}({\bf k}_1-{\bf q},  t_1; {\bf k}_2 t_2) + \qquad
\label{kbeq}
\\\nonumber
\sum\limits_{{\bar {\bf k}}} \Sigma^{{\rm HF}}({\bf k}_1t_1;{\bar {\bf k}}t_1)
G^{\gtrless}({\bar {\bf k}}t_1; {\bf k}_2t_2) +
I^{\gtrless}({\bf k}_1t_1;{\bf k}_2t_2),
\end{eqnarray}
(to be supplemented with the adjoint equation), where $\Sigma^{{\rm HF}}$ is
the Hartree--Fock
selfenergy, and the collision integrals $I^{\gtrless}$ contain the short-range
correlation effects (see below).

As we are interested in the dynamical response of the electron gas to a longitudinal
electrostatic perturbation, we choose
$U({\bf q},t)=U_0(t)\delta_{{\bf q},{\bf q}_0}$.
Before the onset of the field, $t<t_0$, the system is homogeneous,
$G^{\gtrless}({\bf k}_1t_1; {\bf k}_2t_2)\sim \delta_{{\bf k}_1,{\bf k}_2}$,
however, for $t>t_0$, the field gives rise to harmonic modulations
\begin{eqnarray}
G^{\gtrless}_{\mu_1, \mu_2}({\bf k} t_1 t_2) \equiv
G^{\gtrless}({\bf k}+\mu_1 {\bf q}_0, t_1; {\bf k}+\mu_2 {\bf q}_0, t_2),
\label{gmm}
\end{eqnarray}
where $\mu_1$ and $\mu_2$ are integers running  from $-\infty$ to $\infty$.
$q_0$ enters $G_{\mu_1, \mu_2}$ as a parameter and will be omitted.
Due to the symmetry properties
$G^{\gtrless}_{\mu_1+n, \mu_2+n}({\bf k} t_1 t_2) =
G^{\gtrless}_{\mu_1 \mu_2}({\bf k}+n {\bf q}_0, t_1 t_2)$ and
$G^{\gtrless}_{\mu_1, \mu_2}({\bf k} t_1 t_2) = -
G^{\gtrless *}_{\mu_2, \mu_1}({\bf k} t_2 t_1)$, only the matrix elements
$G^{\gtrless}_{n 0}, \; n=0,\pm 1, \pm 2, \dots$ are independent.
As a result, Eq. (\ref{kbeq}) transforms into a system of equations for the functions
$G^{\gtrless}_{n 0}$, i.e.  one of the two momentum arguments of $ G^{\gtrless}$ in
Eq. (\ref{kbeq})
has been replaced by the discrete ``level'' index  $n$. Obviously, this representation
closely resembles the multilevel/multiband kinetic equations (Bloch equations)
familiar from atomic or semiconductor optics, if written in terms of two-time
correlation functions, e.g.  \cite{kwong-etal.98pss}. Only here,
$G_{00}$ corresponds to the spatially homogeneous state, while
$G^{<}_{n0}$ describes transitions of an electron from momentum state
${\bf k}+n {\bf q}_0$
at $t=t_1$ into state ${\bf k}$ at $t=t_2$, cf. definition (\ref{ggl}).
In particular, the equal-time components of $G_{00}^<$ and $G_{m0}^<$ yield
respectively the
homogenous density component $n_0(t)=-i\sum_{{\bf k}} G_{00}^<({\bf k}tt)$
and the field--induced fluctuations
\begin{eqnarray}
\delta n({\bf q},t)&=&\sum_{m\ne 0}\delta n_{m}=
-i\sum_{m\ne 0}\delta_{{\bf q},m {\bf q}_0}\sum_{{\bf k}} G_{m0}^<({\bf k}tt).
\label{deltan}
\end{eqnarray}
In situations where a perturbation treatment of the external field is applicable,
the leading order of the Fourier components of the density is
$\delta n_m \sim O(U_0^m)$. Since the main subject of our paper is the effect of
correlations (collisions) on the plasmon
spectrum, we will focus on the weak-field (linear response) limit below.
Then, we neglect all components of $G^{\gtrless}$ except $G^{\gtrless}_{00}$ and
$G^{\gtrless}_{10}$.
Up to first order in the field, the equations for $G^{\gtrless}_{10}$ read,
for any fixed ${\bf q}_0$ (summation over $m=0,1$ and integration
over ${\bar t}$ from $-\infty$ to $\infty$ is implied),
\begin{eqnarray}
\left( i\hbar \frac{\partial}{\partial t_1} - \epsilon_{{\bf k+q}_0} \right)
  G_{10}^{\gtrless}({\bf k} t_1 t_2) =
 U_0(t_1) \, G_{0 0}^{\gtrless}({\bf k} t_1 t_2) +
\nonumber\\
\Sigma^{{\rm HF}}_{1 m}({\bf k} t_1)
G^{\gtrless}_{m 0}({\bf k} t_1 t_2) +
\nonumber\\
\Sigma_{1 m}^{R}({\bf k}t_1{\bar t})
G_{m 0}^{\gtrless}({\bf k}{\bar t}t_2) +
\Sigma^{\gtrless}_{1 m}({\bf k}t_1{\bar t})
G_{m0}^{A}({\bf k}{\bar t}t_2),
\nonumber\\
\left(- i\hbar \frac{\partial}{\partial t_2} - \epsilon_{{\bf k}} \right)
  G_{10}^{\gtrless}({\bf k} t_1 t_2) =
 U_0(t_2) \, G_{1 1}^{\gtrless}({\bf k} t_1 t_2) +
\nonumber\\
G^{\gtrless}_{1 m}({\bf k} t_1 t_2)
\Sigma^{{\rm HF}}_{m 0}({\bf k} t_2) +
\nonumber\\
G^R_{1 m}({\bf k}t_1{\bar t})
  \Sigma_{m 0}^{\gtrless}({\bf k}{\bar t}t_2) +
G^{\gtrless}_{1 m}({\bf k}t_1{\bar t})
  \Sigma_{m0}^{A}({\bf k}{\bar t}t_2),
\label{g01eq}
\end{eqnarray}
whereas $G^{\gtrless}_{00}$ obey the ``conventional'' spatially homogoneous
field-free equations.
In Eq.~(\ref{g01eq}), $ G_{1 1}^{\gtrless}({\bf k} t_1 t_2)
\equiv G_{0 0}^{\gtrless}({\bf k+q}_0, t_1 t_2)$, the
retarded and advanced functions are defined by ($F$ denotes $G$ or $\Sigma$)
\begin{eqnarray}
F_{\lambda_1\lambda_2}^{R/A}({\bf k}t_1t_2) = \pm \Theta[\pm (t_1-t_2)]\times
\nonumber\\
\left\{ F_{\lambda_1\lambda_2}^{>}({\bf k}t_1t_2)-
F_{\lambda_1\lambda_2}^{<}({\bf k}t_1t_2)\right\}, \quad \lambda_{1,2}=0,1,
\label{fra}
\end{eqnarray}
and the Hartree--Fock selfenergy is
\begin{eqnarray}
\Sigma_{\lambda 0}^{{\rm HF}}({\bf k}t) &=& \delta_{\lambda 1}
V(q_0)\sum\limits_{{\bf p}}
(-i) G_{10}^<({\bf p}tt)
\nonumber\\
&-& \sum\limits_{{\bf p}} (-i) G_{\lambda 0}^<({\bf k-p},tt) V(p),
\quad \lambda=0,1.
\label{shf}
\end{eqnarray}
The selfenergies $\Sigma^{\gtrless}_{10}$ and $\Sigma^{R/A}_{10}$ are of first order
in the field and are obtained from the respective ``00''-components
 by replacing one $G_{00}$ at a time by $G_{10}$ and summing over all
terms generated this way, and $\Sigma_{11}$ follows from $\Sigma_{00}$ by replacing
$G_{00}$ by $G_{11}$.
The equations for $G_{00}$ and $G_{10}$ are to be supplemented by the proper
initial conditions for $G^{\gtrless}_{00}$ corresponding to a correlated spatially
homogeneous electron gas, see below. Furthermore,
$G_{10}({\bf k}t_0t_0)\equiv 0$.

Let us now consider how the dielectric linear response functions can be determined
from the solution of Eqs.~(\ref{g01eq}) and what their properties are.
Using Eq.~(\ref{deltan}), we find in linear response,
\begin{eqnarray}
\sum_{{\bf k}} G_{1 0}^<({\bf k}tt) = i \delta n_{{\bf q}_0}(t) =
\int_{-\infty}^{\infty}d{\bar t} \,
\chi^R({\bf q}_0,t,{\bar t}) U_0({\bar t}),
\label{susc}
\end{eqnarray}
where $\chi^R$ is a retarded susceptibility which,
in general, depends on two times. If the unperturbed system is in a
stationary state,
$\chi^R(t,{\bar t}) \rightarrow \chi^R(t-{\bar t})$,
allowing to apply the convolution theorem to Eq.~(\ref{susc}) with the result
($\omega-$dependence denotes the Fourier component),
\begin{eqnarray}
\chi^{R}({\bf q}_0,\omega)=
\frac{\sum_{{\bf k}} G_{1 0}^<(k,\omega)}{U_0(\omega)},
\label{susc-stat}
\end{eqnarray}
which immediately yields the retarded inverse
dielectric function and the dynamic structure factor
\begin{eqnarray}
\epsilon^{R\,-1}(\omega,{\bf q}_0) &=& 1 + \frac{V(q_0)}{U_0(\omega)}
\sum\limits_{{\bf k}}G_{1 0}^<({\bf k},\omega),
\label{em1}
\\
S(\omega,{\bf q}_0) &=&  -\frac{1}{\pi n_0 U_0(\omega)}
\sum\limits_{{\bf k}}{\rm Im}\,G_{1 0}^<({\bf k},\omega).
\label{s}
\end{eqnarray}

Now, the quality of the plasmon spectrum (\ref{s}) computed from
$G_{10}(t_1 t_2)$ is
fully determined by the approximation for the {\em field-free selfenergies}
$\Sigma_{00}$ in Eqs.~(\ref{g01eq}). In particular, if
$\Sigma_{00}^{{\rm HF}}=\Sigma_{00}^{\gtrless}=0$ (noninteracting electrons
gas), Eq.~(\ref{susc-stat}) reduces to the familiar Lindhard polarization,
$\chi^{R}\equiv\Pi^{R0}$.
Further, if only the Hartree mean field is included,
$\Sigma_{00}^{{\rm HF}}\equiv\Sigma_{00}^{{\rm H}}$,
 one recovers the RPA-BS result (full ring sum),
 $\chi^R=\frac{\Pi^{R0}}{1-V\Pi^{R0}}$,
or, equivalently,
\begin{eqnarray}
\chi^R=\chi^{*}+\chi^{*}V\chi^R,
\label{bse-chi}
\end{eqnarray}
with $\chi^{*} \equiv \Pi^{R0}$. Finally, with the Fock and correlation terms,
$\Sigma^{{\rm F}}$ and $\Sigma_{00}^{\gtrless}$, included also,
one again recovers Eq.~(\ref{bse-chi}), but with a more general expression for
$\chi^{*}$, the proper (irreducible) polarization:
\begin{eqnarray}
\label{chi-star}
\nonumber
\centerline{
\psfig{file=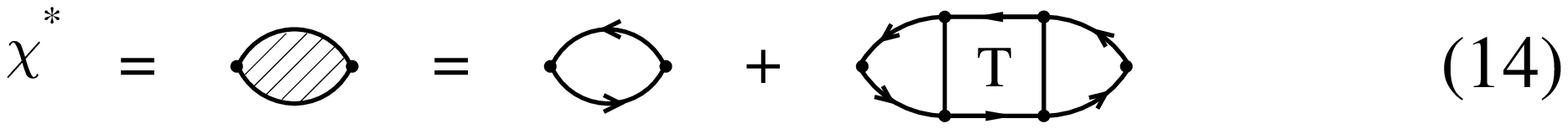,height=0.72cm,width=8.64cm}}
\end{eqnarray}
\begin{eqnarray}\label{teh}
\nonumber
\mbox{with}
\centerline{
\psfig{file=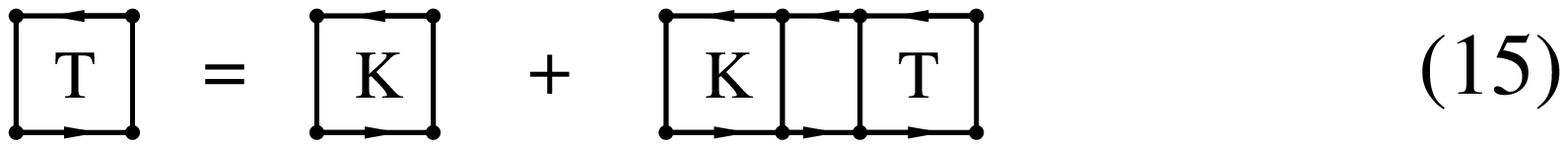,height=0.72cm,width=7.56cm}}
\end{eqnarray}
\addtocounter{equation}{2}

\vspace{-0.5cm}

\noindent
where Eq.~(\ref{chi-star}) starts with $\Pi^{R0}$ (first diagram) but now
contains exchange and correlation corrections (second diagram) in terms of the
particle-hole T-matrix $T$ which obeys the Lippmann--Schwinger equation
(\ref{teh})
with the general interaction kernel $K$ (see below).
One readily recognizes in Eq.~(\ref{bse-chi}) the familiar
{\em field-free Bethe--Salpeter equation} which thus is a direct consequence
of the {\em Kadanoff--Baym equations with weak external field} (\ref{g01eq}).
We underline that
this result applies to equilibrium and {\em arbitrary nonequilibrium} situations
(notice that all derivations are performed on the Keldysh contour, and directed
lines denote {\em full} Green's functions with selfenergy insertions)
\cite{bonitz-etal.99cpp}.

As we demonstrate below, this connection between the BS and KB approaches is
particularly fruitful for the dielectric response of a correlated electron gas:
(i) as the BS approach is a standard formalism for the investigation of
correlation effects, e.g. \cite{green-book}, it can be used to classify
approximations and their properties;
(ii) there exists a one to one correspondence between
the selfenergies $\Sigma_{00}^{\gtrless}$ and the PHI vertex $K$ in
Eq.~(\ref{teh}); (iii) based on the internal consistency of the Kadanoff-Baym
formalism, the properties of the plasmon spectrum are completely determined by
the approximation for $\Sigma_{00}^{\gtrless}$: in particular, density
conservation of $\Sigma_{00}^{\gtrless}$ (which is trivial to meet) guarantees
satisfaction of the f-sum-rule \cite{green-etal.85}.

A valuable practical advantage of the present scheme is that simple
approximations for $\Sigma_{00}^{\gtrless}$
correspond to rather complex approximations for $K$ which allows
for efficient computation of the plasmon spectrum of correlated systems
by solving the KBE (\ref{g01eq}). We demonstrate this below on the example of the
(density conserving) second Born approximation
\begin{eqnarray}
\Sigma^{\gtrless}_{00}({\bf k}tt') = i
\sum_{{\bf p}} |V_{st}({\bf p})|^2 \Pi^{\gtrless}_{00}({\bf k-p},tt')
G_{00}^{\lessgtr}({\bf p},t't)
\label{sborn}
\end{eqnarray}
\vspace{-0.8cm}
\begin{eqnarray}
\label{s-born-d}
\nonumber
\centerline{
\psfig{file=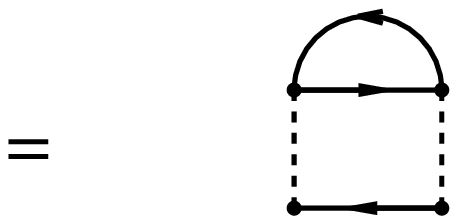,height=1cm,width=9.3cm}}
\end{eqnarray}
where $V_{st}$ is the statically screened Coulomb potential (dashed lines in
the diagram) and
$\Pi^{\gtrless}_{00}$ the
nonequilibrium generalization of the Lindhard polarization bubble,
$\Pi^{\gtrless}_{00}({\bf k}tt')=-i
\sum_{{\bf p}}G^{\gtrless}_{00}({\bf k+p},tt')G^{\lessgtr}_{00}({\bf p}t't)$
Proceeding as in
Ref.~\cite{bonitz-etal.99cpp}, the simple correlation
selfenergy (\ref{sborn}), together with the Fock mean field, transforms into
the following PHI vertex $K$ in the BSE:
\begin{eqnarray}
\label{k-born}
\nonumber
\centerline{
\psfig{file=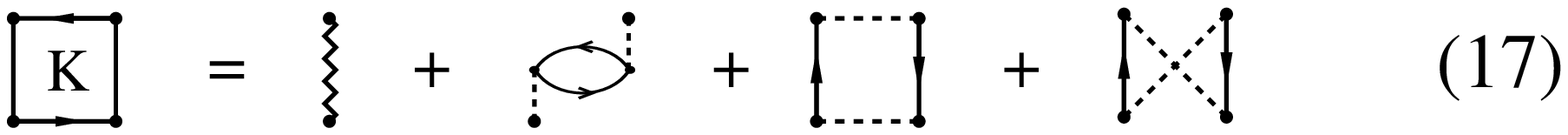,height=0.72cm,width=9.2cm}}
\end{eqnarray}
\addtocounter{equation}{1}

\vspace{-0.3cm}

\noindent
and contains contributions from particle-hole (unscreened) Coulomb
scattering (first diagram, zig-zag line denotes bare Coulomb potential $V$),
excitation of a particle-hole pair (second diagram) and
scattering between two particle-hole pairs (last two diagrams).
Comparison of Eqs.~(\ref{sborn}) and
(\ref{k-born}) reveals the familiar relation between $\Sigma$ and $K$:
$K=V+\delta \Sigma_{00}/\delta G$ \cite{baym-etal.61,green-book}.

While it is very difficult to solve the BSE with kernel~(\ref{k-born})
without further simplifying approximations,
solving Eq.~(\ref{g01eq}) with the selfenergy Eq.~(\ref{sborn}) and the
$\Sigma_{10}$ derived from it is quite straightforward.
We performed numerical solutions for a strongly correlated electron gas in
equilibrium using the numerical procedure which was developed before for the
two--time semiconductor Bloch equations \cite{kwong-etal.98pss}. To create a
correlated initial state,  we run the field-free program for a time longer than
the correlation time starting from an uncorrelated distribution. After this,
the field $U$ was turned on, where we chose a
pulse shape for $U_0(t)$ broad enough to cover the plasmon spectrum.
The thus excited density fluctuation is shown in Fig.~1 for a 3D electron gas
with Brueckner parameter $r_s=4$ and temperature $k_BT=0.69 E_F$ (Fermi energy)
for two wavenumbers. For comparison, also the
uncorrelated result is shown (Hartree--Fock selfenergies only which is
equivalent to RPA-BS plus exchange).
While $\delta n(t)$ depends on the explicit form of $U_0(t)$, obviously
the linear response quantities $\epsilon^R(\omega)$ and $S(\omega)$,
Eqs.~(\ref{em1}, \ref{s}), are independent of $U_0$.
Fig.~2 shows the dynamic structure factors, corresponding
to the results in Fig.~1.
Clearly, one sees that the short-range correlations lead to a damping of
$\delta n(t)$ in excess of the collisionless Landau damping, cf. Fig.~1,
which corresponds to a red shift and an additional broadening of the plasmon
peak in the structure factor, Fig.~2. Remarkably, our numerical scheme
preserves the f-sum rule for the small (large) wave number to $0.03$\%
($0.6$\%). In contrast, neglecting terms in $K$ but keeping the
selfenergy insertions in G leads to violation of the sum
rule. For example, for inclusion of the first two diagrams only
(curve ``1+2'' in the inset of Fig.~2) and the first diagram only (``1'')
the corresponding numbers are, respectively, $2.1$\%  ($0.8$\%) and
$1346$\%  ($416$\%), and for still smaller $q_0$ the error increases
rapidly \cite{tamme-etal.99}. Clearly, the first two diagrams are the most
important ones in producing the overall features of $S(\omega)$, but, as the
inset of Fig.~2 shows, the last two diagrams may still be needed to give the
correct high-energy behavior.

In summary, we have presented a new selfconsistent approach to the dielectric
properties of a correlated electron gas. Using an ``interband'' generalization
and solving the problem in the time domain allows to take maximum advantage
of the selfconsistent Kadanoff-Baym scheme:
simple approximations for the collision integrals transform into complex
correlation corrections in the plasmon spectrum with ``built in'' sum rule
preservation. This scheme is straightforwardly extendable to higher order
correlations. Moreover, it applies to arbitrary {\em nonequilibrium}
situations, and is easily generalized to the {\em nonlinear} dielectric
response in strong fields.

This work is supported by the Deutsche Forschungsgemeinschaft
(Schwerpunkt ``Quantenkoh\"arenz in Halbleitern'').  We acknowledge
useful discussions with D. Kremp, R. Schepe, and S. K\"ohler.

\begin{figure}
\caption{Density fluctuation of a strongly correlated electron gas with
 for two wavenumbers. For comparison, the
uncorrelated response for one wavenumber (dotted line) and the exciting
field (dashes) are shown too. $k_F$ denotes the Fermi momentum, Ry$=13.6 eV$.}
\end{figure}
\begin{figure}
\caption{Dynamic structure factor (\ref{s}) for the correlated electron gas
of Fig.~1 (same line styles). Inset shows $S$ for $q_0=0.62 k_F$ and contains
two other approximations to the correlations corresponding to retaining the
first diagram in Eq. (17) and first plus second diagrams, respectively.}
\end{figure}


\begin{references}

\bibitem{schoene-etal.98} W.D. Sch\"one, and A.G. Eguiluz,
                Phys. Rev. Lett. {\bf 81}, 1662 (1998)
\bibitem{schindlmayr-etal.98} A. Schindlmayr, and R.W. Godby,
                Phys. Rev. Lett. {\bf 80}, 1702 (1998)
\bibitem{tamme-etal.99} D. Tamme, R. Schepe, and K. Henneberger,
                Comment on \cite{schoene-etal.98},
                Phys. Rev. Lett. (1999)
\bibitem{green-etal.85} F. Green, D. Neilson, and J. Szymanski,
               Phys. Rev. B {\bf 31}, 2779 (1985)
\bibitem{theobald-etal.96} W. Theobald, R. H\"assner, C. W\"ulker, and
         R. Sauerbrey, Phys. Rev. Lett. {\bf 77}, 298 (1996)
\bibitem{wierling-etal.} G. R\"opke, and A. Wierling,
         Phys. Rev. E {\bf 57}, 7075 (1998)
\bibitem{green-book} W.D. Kraeft, D. Kremp, W. Ebeling, and G. R\"opke,
               {\em Quantum Statistics of Charged Particle Systems},
               Akademie-Verlag Berlin, 1986
\bibitem{kalman-etal.99} G. Kalman, V. Valtchinov, and K.I. Golden,
         Phys. Rev. Lett. {\bf 82}, 3124 (1999)
\bibitem{mermin70} D. Mermin, Phys. Rev. B {\bf 1} (1970)
\bibitem{baym-etal.61} G. Baym, and L.P. Kadanoff,
               Phys. Rev. {\bf 124}, 287 (1961)
\bibitem{green-etal.87} F. Green, D. Neilson, D. Pines, and J. Szymanski,
               Phys. Rev. B {\bf 35}, 133 (1987)
\bibitem{farinas-etal.99} P.F. Farinas, K.S. Bedell, and N. Stuart,
Phys. Rev. Lett. {\bf 82}, 3851 (1999)
\bibitem{dang-etal.98} N.D. Dang, and A. Arima,
Phys. Rev. Lett. {\bf 80}, 4145 (1998)
\bibitem{bonitz-etal.99cpp} M. Bonitz, N.H. Kwong, D. Semkat, and D. Kremp,
                Contrib. Plasma Phys. {\bf 39}, 37 (1999)
\bibitem{schaefer-etal.86} W. Sch\"afer, and J. Treusch,
              Z. Physik B {\bf 63}, 407 (1986)
\bibitem{kwong-etal.98pss} N.H. Kwong, M. Bonitz, R. Binder, and H.S. K\"ohler,
              phys. stat. sol. (b) {\bf 206}, 197 (1998)
\end{references}
\end{document}